\documentclass{ws-ijmpe}

\begin{document}

\markboth{Debasish Das}{Identical Meson Interferometry in STAR Experiment}

\catchline{}{}{}{}{}

\title{Identical Meson Interferometry in STAR Experiment}


\author{\footnotesize Debasish Das  ( for the STAR Collaboration )
\footnote{http://www.star.bnl.gov}}

\address{Variable Energy Cyclotron Centre, 1/AF, Bidhan Nagar\\
Kolkata - 700064, India.\\
ddas@veccal.ernet.in, debasish@rcf.rhic.bnl.gov}



\maketitle

\begin{history}
\end{history}

\begin{abstract}

The influence of Bose-Einstein statistics on multi-particle
production characterized for various systems and energies by the
STAR collaboration provides interesting information about the 
space-time dynamics of relativistic heavy-ion collisions at RHIC.
We present the centrality and transverse mass dependence measurements 
of the two-pion interferometry in Au+Au collisions
at $\sqrt{s_{\rm{NN}}}$ = 62.4 GeV and Cu+Cu collisions at 
$\sqrt{s_{\rm{NN}}}$ = 62.4 GeV and 200 GeV. We compare the new data
with previous STAR measurements from p+p, d+Au and Au+Au collisions
at $\sqrt{s_{\rm{NN}}}$ = 200 GeV. In all systems and centralities, 
HBT radii decrease with transverse mass in a similar manner, which is 
qualitatively consistent with collective flow.
The scaling of the apparent freeze-out volume 
with the number of participants and charged particle multiplicity 
is studied. Measurements of Au+Au collisions at same centralities 
and different energies yield different freeze-out volumes, which 
mean that $N_{part}$ is not a suitable scaling variable. 
 The multiplicity scaling of the measured HBT radii is found to be
independent of colliding system and collision energy.

\end{abstract}

\section{Introduction}

The information about the space-time structure of the emitting source created in 
elementary particle and heavy ion collisions from the measured particle momenta can 
be extracted by the method of two-particle intensity interferometry techniques 
also called the $\it {Hanbury-Brown - Twiss (HBT) effect}$~\cite{Goldhaber:1960sf,Wiedemann:1999qn} 
which was initially developed to measure the angular size of distant stars~\cite{HanburyBrown:1954wr}.
HBT is a useful method to understand the crucial reaction mechanisms of the particle
emitting source in relativistic heavy ion collisions, where the Quark-Gluon Plasma(QGP) 
is expected to be formed. Experimental search for evidence of QGP is the 
most fundamental and challenging task in modern nuclear physics.

In this paper, we present the comparative pion intensity interferometry measurements 
from large collision systems of Au+Au and Cu+Cu at $\sqrt{s_{\rm{NN}}}$ = 62.4 and 200 GeV 
with small systems like p+p and d+Au at $\sqrt{s_{\rm{NN}}}$ = 200 GeV using the Solenoidal
Tracker At RHIC(STAR) detector at Relativistic Heavy Ion Collider(RHIC). The wealth of
new data from STAR detector with its large acceptance and full azimuth capabilities have
provided a new systematic study of the hot and dense medium created in ultra-relativistic
heavy-ion collisions. The measured HBT radii are studied as a function of average 
transverse momentum ($k_{T}$ $=$ ($|\overrightarrow{p_{1}}_{T}$ $+$ $\overrightarrow{p_{2}}_{T}|$)$/$2) 
in 4 bins that correspond to [150,250] MeV/c, [250,350] MeV/c, [350,450] MeV/c and [450,600] MeV/c. 
The results are presented and discussed as a function of average $k_{T}$ 
( or $m_{T}$ = $\sqrt{k_{T}^{2}+m_{\pi}^{2}}$ ) in each of those bins.
The $m_{T}$ dependence of HBT radii for all RHIC energies in Au+Au and Cu+Cu collisions 
are compared for the first time.

\section{$m_T$ and energy dependence of HBT parameters in Au+Au and Cu+Cu collisions}

\begin{figure}[th]
\centerline{\psfig{file=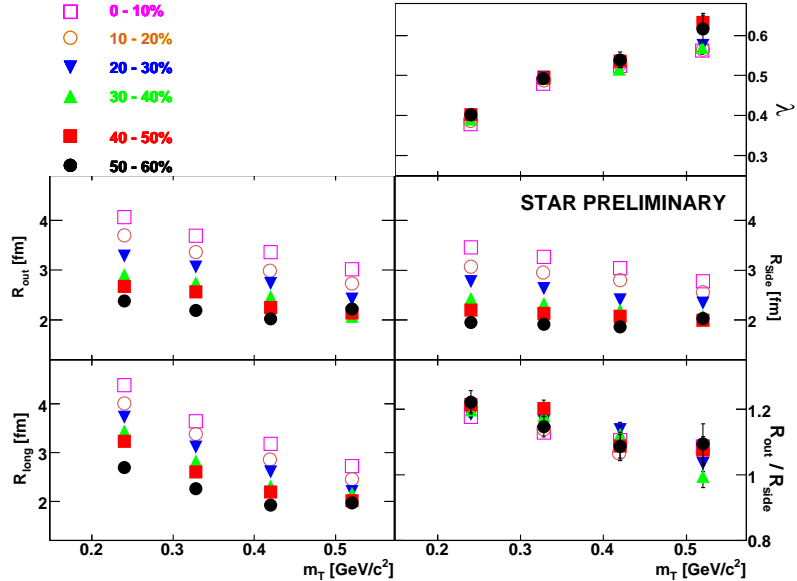,width=11cm}}
\vspace*{8pt}
\caption{\label{fig:Fig1}The HBT parameters vs $m_{T}$ for 6 different centralities for Cu+Cu 
collisions at $\sqrt{s_{\rm{NN}}}$ = 62.4 GeV. Only statistical errors are shown.}
\end{figure}

The first measured centrality dependence of HBT radius parameters for Cu+Cu collisions at 
$\sqrt{s_{\rm{NN}}}$ = 62.4 GeV is shown in Fig.~\ref{fig:Fig1}. The three HBT radii 
$R_{out}$, $R_{side}$ and $R_{long}$ increase with increasing centrality. The $R_{out}$$/$$R_{side}$ 
ratio $\sim$ 1 and exhibits no clear centrality dependence.

As shown in Fig.~\ref{fig:Fig1} the decrease of HBT radii with $m_T$ at all observed 
centralities is qualitatively consistent with collective 
flow~\cite{Pratt:1984su,Kolb:2003dz,Retiere:2003kf,Hirano:2002hv}. 
 The presence of collective flow in the expanding system causes a decrease 
in ``HBT radii'' with $m_{T}$~\cite{Lisa:2005dd,Schlei:1996mc} 
where the fall-off of ``out'' and ``side'' components is caused by
transverse flow~\cite{Adams:2004yc,Heinz:1996bs,Wiedemann:1997cr}  
and for the ``long'' component due to the longitudinal 
flow~\cite{Adams:2004yc,Wiedemann:1997cr}. The $\lambda$ parameter increases 
with $m_{T}$ as observed in previous STAR measurements in Au+Au collisions at 
$\sqrt{s_{\rm{NN}}}$ = 200 GeV~\cite{Adams:2004yc}. Such increase is due to the
decreasing contribution of pions produced from long-lived resonance decays at higher transverse
momenta. 

\begin{figure}[th]
\centerline{\psfig{file=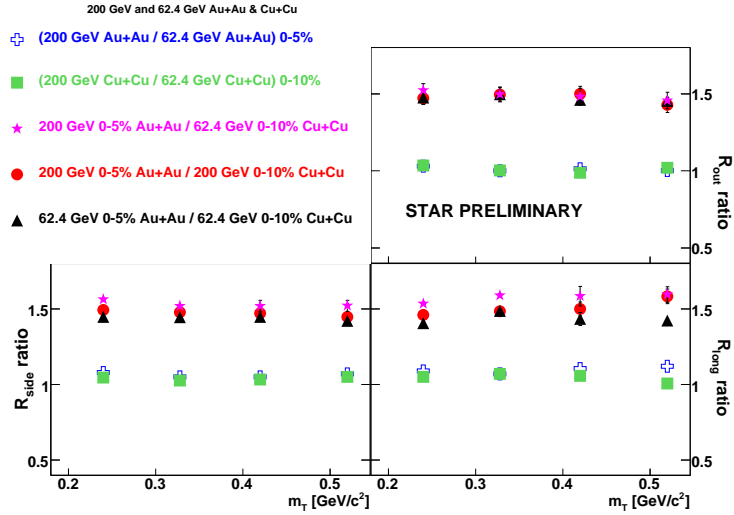,width=10cm}}
\vspace*{8pt}
\caption{\label{fig:Fig2} The HBT radii ratio at top centralities for Au+Au and Cu+Cu collisions.}
\end{figure}

The ratios of transverse and longitudinal HBT radii at top centralities in Au+Au and Cu+Cu collisions
at $\sqrt{s_{\rm{NN}}}$ = 200 GeV and 62.4 GeV is presented in Fig.~\ref{fig:Fig2}, from which we infer
that the corresponding HBT radii vary similarly with $m_{T}$. 

\section{Volume estimates and multiplicity scaling}

The pion freeze-out volume estimates using the HBT radii will provide an understanding
of the freeze-out properties. The freeze-out volume can be estimated using the expressions like :
\begin{subequations}
\begin{equation}
\label{eq:one}
V_{f} \propto R_{side}^2R_{long}
\end{equation}
\begin{eqnarray}
\label{eq:two}
~~~~~~\propto R_{out}R_{side}R_{long}
\end{eqnarray}
\end{subequations}
where $V_{f}$ is the freeze-out volume.

 We have discussed in the previous section that the effects 
of collective expansion of the system lead to  the $m_{T}$ dependence fall-off on HBT 
radius parameters compared to the measured dimensions of source. The femtoscopic 
radii of such an expanding source correspond to the $m_{T}$(or $k_{T}$) dependent region 
of homogeneity which is smaller than the entire collision region~\cite{Wiedemann:1999qn,Lisa:2005dd}. 
Henceforth the best volume estimates in Eq.(\ref{eq:one} and \ref{eq:two}) are measured for lowest 
$k_{T}$ bin which in our case corresponds to [150,250]MeV/c as described previously.

\begin{figure}[th]
\centerline{\psfig{file=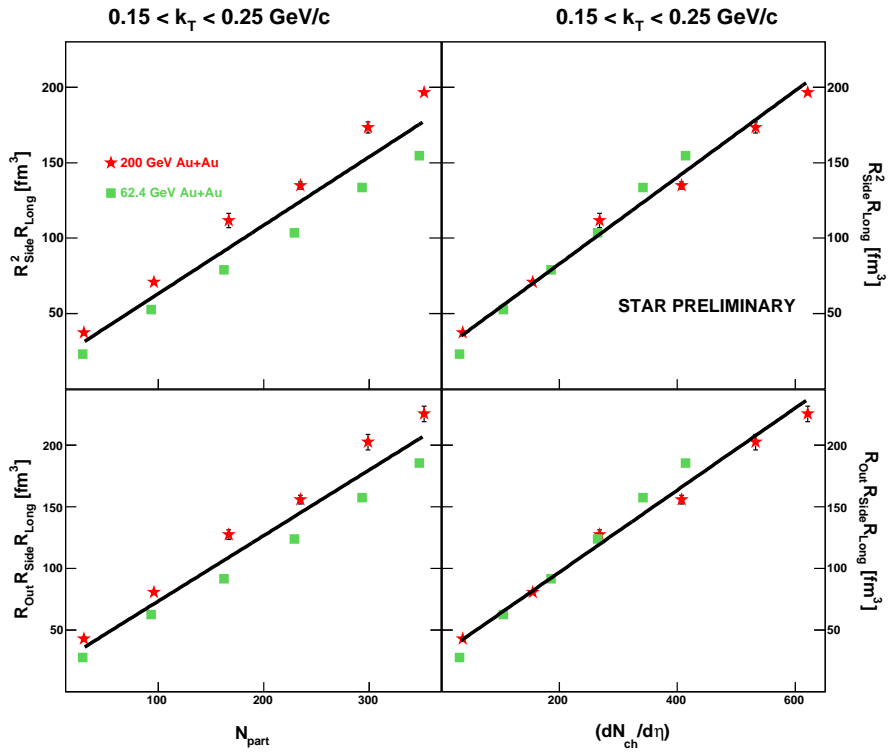,width=12cm}}
\vspace*{8pt}
\caption{\label{fig:Fig4}Pion freeze-out volume estimates as a function of number of 
participants and charged particle multiplicities for Au+Au collisions at $\sqrt{s_{\rm{NN}}}$ = 200 and 
62.4 GeV. The lines are plotted to guide the eye and represent linear fits to data.}
\end{figure}

Fig.~\ref{fig:Fig4} shows the comparative study of freeze-out volume estimates
(using Eq.(\ref{eq:one}) and Eq.(\ref{eq:two})) as a function of the number of 
participants and charge particle multiplicity for Au+Au collisions, where the earlier
measurements at  $\sqrt{s_{\rm{NN}}}$ = 200 GeV~\cite{Adams:2004yc} are presented
with the preliminary STAR results from 62.4 GeV. Measurements of Au+Au collisions at 
same centralities and different energies show different freeze-out volume, 
which means that $N_{part}$(initial overlap geometry) is not a suitable 
scaling variable in this case. On the other hand charge particle 
multiplicity(final freeze-out geometry) seems to be a better scaling variable~\cite{Lisa:2005dd}.

\begin{figure}[th]
\centerline{\psfig{file=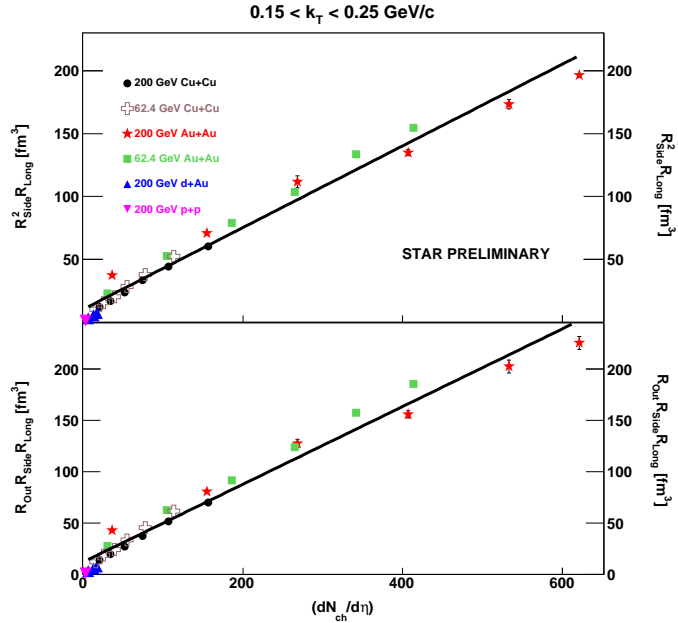,width=10cm}}
\vspace*{8pt}
\caption{\label{fig:Fig5}Pion freeze-out volume estimates as a function of charged particle multiplicity. 
The lines are plotted to guide the eye and represent linear fits to data.}
\end{figure}

\begin{figure}[th]
\centerline{\psfig{file=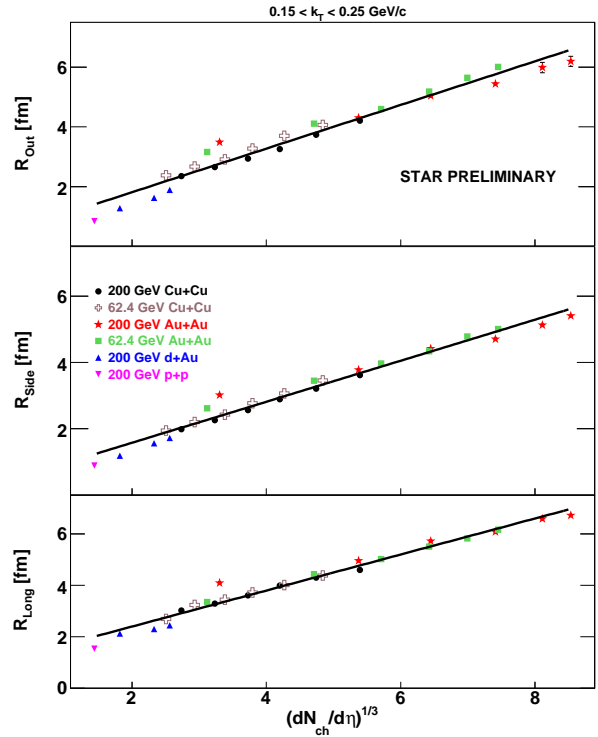,width=8cm}}
\vspace*{8pt}
\caption{\label{fig:Fig6}Pion source radii dependence on charged particle multiplicity. 
The lines are plotted to guide the eye and represent linear fits to data.}
\end{figure}

The study of freeze-out volume estimates are presented in Fig.~\ref{fig:Fig5} for 
p+p, d+Au~\cite{Chajecki:2005zw}, Cu+Cu and Au+Au collisions as a function 
of charge particle multiplicity. The freeze-out volume estimates for measured systems 
show a linear dependence as a function of charge particle multiplicity. The linear 
dependence of HBT radii with $(dN_{ch}/d\eta)^{1/3}$  for all STAR 
data-sets is exhibited in Fig.~\ref{fig:Fig6}. Such 
dependences~\cite{Lisa:2005dd} are naturally expected within a framework of universal mean-free-path of 
pions at freeze-out.~\cite{Adamova:2002ff}. 

\section{Conclusions}

The systematic study of HBT radii for varied initial energy and system are presented.
For the systems studied, the multiplicity and $m_{T}$(or $k_{T}$) dependence of the
radii are strongly consistent with the previously measured colliding systems at RHIC.  
Measurements of Au+Au collisions at same centralities 
and different energies yield different freeze-out volumes, which 
infer that $N_{part}$ is not a suitable scaling variable. The radii scale 
with the collision multiplicity; in a static model, this is consistent 
with a universal mean-free-path at freeze-out.  As in measurements at all 
other energies~\cite{Lisa:2005dd,Chajecki:2005zw}, the $m_{T}$(or $k_{T}$) 
dependence remains independent of $\sqrt{s_{\rm{NN}}}$, collision system, and multiplicity.

\end{document}